\UseRawInputEncoding
\documentclass[12pt,letter]{article}

\usepackage{subfigure}
\usepackage{graphicx, epsfig, color}
\usepackage{amsmath,amssymb}
\def\eslt{\not\!\!{E_T}}

\def\bi{\begin{itemize}}
\def\ei{\end{itemize}}

\def\tchi{\tilde\chi}

\def\sps1ap{SPS1a$^\prime$}
\def\c1p{C1$^\prime$}

\def\tst{\tilde t}

\def\tg{\tilde g}

\def\alt{\stackrel{<}{\sim}}
\def\agt{\stackrel{>}{\sim}}
\def\be{\begin{equation}}  
\def\ee{\end{equation}}  
\def\bea{\begin{eqnarray}}  
\def\eea{\end{eqnarray}}  
\def\beas{\begin{eqnarray*}}  
\def\eeas{\end{eqnarray*}}

\begin{document}
\begin{titlepage}
\begin{flushright}
OU-HEP-210403
\end{flushright}

\vspace{0.5cm}
\begin{center}
{\Large \bf Distribution of supersymmetry $\mu$ parameter\\
and Peccei-Quinn scale $f_a$ from the landscape
}\\ 
\vspace{1.2cm} \renewcommand{\thefootnote}{\fnsymbol{footnote}}
{\large Howard Baer$^{1}$\footnote[1]{Email: baer@nhn.ou.edu },
Vernon Barger$^{2}$\footnote[2]{Email: barger@pheno.wisc.edu },
Dibyashree Sengupta$^{3}$\footnote[3]{Email: Dibyashree.Sengupta-1@ou.edu}\\
and Robert Wiley Deal$^{1}$\footnote[3]{Email: rwileydeal@ou.edu}
}\\ 
\vspace{1.2cm} \renewcommand{\thefootnote}{\arabic{footnote}}
{\it 
$^1$Homer L. Dodge Dep't of Physics and Astronomy,
University of Oklahoma, Norman, OK 73019, USA \\
}
{\it 
$^2$Dep't of Physics,
University of Wisconsin, Madison, WI 53706, USA \\
}
{\it 
$^3$Dep't of Physics,
National Taiwan University, Taipei 10617, Taiwan, R. O. C. \\
}
\end{center}

\vspace{0.5cm}
\begin{abstract}
\noindent 
A scan of soft SUSY breaking parameters within the string theory landscape
with the MSSM assumed as the low energy effective field theory-- 
using a power-law draw to large soft terms coupled with an anthropic selection 
of a derived weak scale to be within a factor four of our measured value-- 
predicts a peak probability of $m_h\simeq 125$ GeV with sparticles masses
typically beyond the reach of LHC Run 2. 
Such multiverse simulations usually assume a fixed value of the 
SUSY conserving superpotential $\mu$ parameter to be 
within the assumed anthropic range, $\mu\alt 350$ GeV. 
However, depending on the assumed solution to the SUSY $\mu$ problem, 
the expected $\mu$ term distribution can actually be derived. 
In this paper, we examine two solutions to the SUSY $\mu$ problem. 
The first is the gravity-safe-Peccei-Quinn (GSPQ) model based on an assumed 
$\mathbb{Z}_{24}^R$ discrete $R$-symmetry which allows a gravity-safe
accidental, approximate Peccei-Quinn global symmetry to emerge which
also solves the strong CP problem. The second case is the Giudice-Masiero
solution wherein the $\mu$ term effectively acts as a soft term 
and has a linear draw to large values. For the first case, we also
present the expected landscape distribution for the PQ scale $f_a$; 
in this case, weak scale anthropics limits its range to 
the cosmological sweet zone of around $f_a\sim 10^{11}$ GeV.
\vspace*{0.8cm}


\end{abstract}

\end{titlepage}

\section{Introduction}
\label{sec:intro}

One of the curiosities of nature pertains to the origin of mass scales.
Naively, one might expect all mass scales to be of order the fundamental
Planck mass scale $m_{Pl}=1.2\times 10^{19}$ GeV as occurs in 
quantum mechanics and in its relativistic setting: string theory.
For instance, one expects the cosmological constant $\Lambda_{cc}\sim m_{Pl}^2$
whereas its measured value is over 120 orders of magnitude smaller.
The only plausible explanation so far is by Weinberg\cite{Weinberg:1987dv} 
in the context of the eternally inflating multiverse wherein each pocket
universe has a different value of $\Lambda_{cc}$ ranging from 
$-m_{Pl}^2$ to $+m_{Pl}^2$: if $\Lambda_{cc}$ were too much larger 
than its measured value, then the early universe would have expanded so quickly
that structure in the form of galaxies, and hence observors,  
would not occur. This anthropic explanation finds a natural setting in 
the string theory landscape of vacuum solutions\cite{Bousso:2000xa} 
where of order $10^{500}$\cite{Denef:2004ze} (or many, many more\cite{Taylor:2015xtz}) solutions 
may be expected from string flux compactifications\cite{Douglas:2006es}.

A further mystery is the origin of the weak scale: why is 
$m_{weak}\sim m_{W,Z,h}\sim 100$ GeV instead of $10^{19}$ GeV?
A similar environmental solution has been advocated by Agrawal, Barr,
Donoghue and Seckel (ABDS)\cite{Agrawal:1997gf,Agrawal:1997gf2}:  
if $m_{weak}$ was a factor $2-5$ greater than its measured value, 
then quark mass differences would be affected such that complex nuclei, 
and hence atoms as we know them, could not form (atomic principle).

This latter solution has been successfully applied in the context of weak scale
supersymmetry (WSS)\cite{WSS} within the string theory landscape. 
The assumption here is to adopt a fertile patch of landscape vacua
where the Minimal Supersymmetric Standard Model forms the 
correct weak scale effective field theory (EFT), but wherein the soft 
SUSY breaking terms would scan in the landscape. For perturbative SUSY
breaking where no non-zero $F$-term or $D$-term is favored over any other 
in the landscape, then soft terms are expected to scan as a 
power-law\cite{Susskind:2004uv,Douglas:2004qg,ArkaniHamed:2005yv}:
\be
f_{SUSY}\sim m_{soft}^n
\label{eq:fsusy}
\ee
where $n=2n_F+n_D-1$ with $n_F$ the number SUSY breaking hidden sector 
$F$-terms and $n_D$ is the number of SUSY breaking hidden sector $D$-terms.
The factor two comes from the fact that $F$-terms are distributed as complex
values whilst the $D$-breaking fields are distributed as real numbers.
Even for the textbook value $n_F=1$ and $n_D=0$, already one expects
a statistical draw from the landscape to large soft SUSY breaking terms
and one might expect soft terms at the highest possible scale, perhaps
at the Planck scale.

However, such huge soft terms would generically result in a Higgs potential
with either charge-or-color breaking minima (CCB) or no electroweak symmetry
breaking (EWSB) at all. For vacua with appropriate EWSB, then one typically
expects the pocket universe value of the weak scale 
$m_{weak}^{PU}\gg m_{weak}^{OU}$ in violation of the atomic principle
(where $m_{weak}^{OU}$ corresponds to the measured value of the weak 
scale in our universe).
Here, for specificity, we will evaluate the expected weak scale value in
terms of $m_Z^{PU}$ as calculated for each pocket universe via the 
SUSY EWSB minimization conditions, which read
\bea 
\frac{(m_Z^{PU})^2}{2}& =& \frac{m_{H_d}^2 + \Sigma_d^d -(m_{H_u}^2+\Sigma_u^u)\tan^2\beta}{\tan^2\beta -1} -\mu^2\\ 
&\simeq & -m_{H_u}^2-\Sigma_u^u(\tst_{1,2})-\mu^2 .
\label{eq:mzs}
\eea 
Here, $m_{H_u}^2$ and $m_{H_d}^2$ are squared soft SUSY breaking
Lagrangian terms, $\mu$ is the superpotential Higgsino mass parameter,
$\tan\beta =v_u/v_d$ is the ratio of Higgs field
vacuum-expectation-values (vevs) and the $\Sigma_u^u$ and $\Sigma_d^d$
contain an assortment of radiative corrections, the largest of which
typically arise from the top squarks.  Expressions for the $\Sigma_u^u$
and $\Sigma_d^d$ are given in the Appendix of Ref. \cite{rns}. 

To remain in accord with the atomic principle according to Ref. \cite{Agrawal:1997gf,Agrawal:1997gf2}, 
we will require, for a derived value of $\mu$ (so that $\mu$ is not available
for the usual finetuning in Eq. \ref{eq:mzs} needed to gain the 
measured value of $m_Z^{OU}$), 
that $m_Z^{PU}<4 m_Z^{OU}$ where $m_Z^{OU}=91.2$ GeV. 
This constraint is then the same as requiring the electroweak naturalness
parameter\cite{ltr,rns} $\Delta_{EW}\alt 30$. 
Thus, the anthropic condition is that
-- for various soft term values selected statistically according to 
Eq. \ref{eq:fsusy}-- there must be appropriate EWSB 
(no CCB or non-EWSB vacua) and that $m_{Z}^{PU}<4 m_Z^{OU}$. 
These selection requirements have met with success within the 
framework of gravity-mediation (NUHM2) models\cite{Baer:2017uvn} 
and mirage mediation\cite{Baer:2019tee} (MM) 
in that the probability distribution for the Higgs mass $m_h$ ends up 
with a peak around $m_h\sim 125$ GeV with sparticle masses typically 
well beyond LHC limits. Such results are obtained for $n=1,\ 2,\ 3$ and $4$
and even for a $\log(m_{soft})$ distribution\cite{Broeckel:2020fdz,Baer:2020dri}.

These encouraging results were typically obtained by fixing the
SUSY conserving $\mu$ parameter at some natural value 
$\mu\alt 4m_Z^{OU}\sim 350$ GeV so that the atomic principle isn't 
immediately violated. But what sort of distribution of SUSY $\mu$ parameter
is expected from the landscape? The answer depends on what sort of 
solution to the SUSY $\mu$ problem is assumed in the underlying model
(a recent review of 20 solutions to the SUSY $\mu$ problem
is given in Ref. \cite{Bae:2019dgg}). Recall that since $\mu$ is SUSY
conserving and not SUSY breaking, then one might expects its value 
to be far higher than $m_{weak}$, perhaps as high as the reduced 
Planck mass $m_{P}$. 
But phenomenologically, its value ought to be at or around the weak scale
in order to accommodate appropriate EWSB\cite{Polonsky:1999qd}.

In this paper, our goal is to calculate the expected $\mu$ parameter
probability distribution expected from the string landscape from two
compelling solutions to the SUSY $\mu$ problem. 
We will first examine the so-called gravity-safe Peccei-Quinn (GSPQ) model\footnote{
The GSPQ model\cite{Baer:2018avn} is a hybrid between the CCK\cite{Choi:1996vz} 
and BGW\cite{bgw} models.} which is based upon a discrete 
$R$-symmetry $\mathbb{Z}_{24}^R$ 
from which the global PQ emerges as an accidental, approximate symmetry; 
it then solves the SUSY $\mu$ problem and the strong CP problem 
in a gravity-safe manner\cite{Baer:2018avn}.
The second solution is perhaps most popular: 
the Giudice-Masiero (GM) mechanism\cite{Giudice:1988yz} wherein the 
$\mu$ parameter arises from non-renormalizable terms in the K\"ahler
potential. 

\section{Distribution of $\mu$ parameter and PQ scale 
for the GSPQ model}
\label{sec:GSPQ}

The first $\mu$ term solution we will examine is the so-called
gravity-safe PQ (GSPQ) model which was specified in Ref. \cite{Baer:2018avn}.
The GSPQ model is based upon a discrete $\mathbb{Z}_{24}^R$ $R$-symmetry 
to at first forbid the $\mu$ parameter. 
The set of discrete $R$ symmetries that allow for all anomaly-cancellations 
in the MSSM (up to Green-Schwarz terms) 
and are consistent with $SO(10)$ or $SU(5)$ GUT matter assignments 
were catalogued by Lee {\it et al.} in Ref. \cite{Lee:2011dya} and found to 
consist of $\mathbb{Z}_4^R$, $\mathbb{Z}_6^R$, $\mathbb{Z}_8^R$,
$\mathbb{Z}_{12}^R$ and $\mathbb{Z}_{24}^R$. These discrete $R$-symmetries
1. forbid the SUSY $\mu$ term, 2. forbid all $R$-parity-violating operators,
3. suppress dimension-5 proton decay operators while 4. allowing
for the usual superpotential Yukawa and neutrino mass terms.

The superpotential for the GSPQ model introduces two additional
PQ sector fields $X$ and $Y$ and is given by
\bea
W_{GSPQ}&=& f_u QH_uU^c+f_dQH_dD^c+f_\ell LH_dE^c +f_\nu L H_u N^c
+M_N N^c N^c/2\nonumber \\
& +&\lambda_\mu X^2 H_u H_d/m_P +f X^3 Y/m_P , 
\label{eq:W_GSPQ}
\eea
where $f_{u,d,\ell ,\nu}$ are the usual MSSM+right-hand-neutrino (RHN)
Yukawa couplings and $M_N$ is a Majorana neutrino mass term which is essential 
for the SUSY neutrino see-saw mechanism. 
Since the $\mu$ term arises from the PQ sector of the superpotential 
(second line of Eq. \ref{eq:W_GSPQ}), this is an example of the Kim-Nilles
solution to the SUSY $\mu$ problem\cite{KN}. The GSPQ model is 
a hybrid between the Choi-Chun-Kim\cite{Choi:1996vz} (CCK) 
radiative PQ breaking model 
and the Babu-Gogoladze-Wang model\cite{bgw} (BGW) 
based on discrete gauge symmetries. 
For the case of $\mathbb{Z}_{24}^R$ symmetry applied to the GSPQ model, 
then it was also found that all further non-renormalizable contributions
to $W_{GSPQ}$ are suppressed by powers up to $1/m_P^7$: 
terms such as $X^8Y^2/m_P^7$ and $X^4Y^6/m_P^7$ being allowed. 
These terms contribute to the scalar potential with terms
suppressed by powers of $1/m_P^8$. The wonderful result is that
the Peccei-Quinn symmetry needed to resolve the strong CP problem
emerges as an accidental, approximate symmetry much like baryon- and 
lepton-number emerge in the SM as a result of the SM gauge symmetries.
The $\mathbb{Z}_{24}^R$ symmetry is strong enough to sufficiently 
suppress PQ breaking terms in $W_{GSPQ}$ such that a very sharp
PQ symmetry emerges: enough to guarantee that PQ-violating
contributions to the strong CP violating $\bar{\theta}$ parameter
keep its value below $\bar{\theta}\alt 10^{-10}$ in accord with 
neutron EDM measurements. Thus, the GSPQ model based on 
$\mathbb{Z}_{24}^R$ discrete $R$-symmetry yields a 
{\it gravity-safe} global PQ symmetry!

The PQ symmetry ends up being violated when SUSY breaking also breaks 
the $\mathbb{Z}_{24}^R$ discrete $R$-symmetry, leading to the emergence
of the $\mu$ parameter with value $\mu\sim \lambda_{\mu} v_X^2/m_P$. 
In the GSPQ model, the $F$-term part of the scalar potential
\be
V_F=|3f\phi_X^2\phi_Y/m_P|^2+|f \phi_X^3/m_P|^2
\ee
is augmented by SUSY breaking soft term contributions
\be
V_{soft}\ni m_X^2|\phi_X|^2+m_Y^2|\phi_Y|^2+(f A_f\phi_X^3\phi_Y/m_P +h.c.).
\ee
SUSY breaking with a large value of trilinear soft term $-A_f$ 
leads to $\mathbb{Z}_{24}^R$ breaking (allowing a $\mu$ term to develop)
and consequent breaking of the approximate, accidental PQ symmetry, 
leading to the pseudo-Goldstone boson axion $a$ (a combination of the
$X$ and $Y$ fields).

The GSPQ scalar potential minimization conditions are\cite{Bae:2014yta} 
(neglecting the Higgs field contributions which lead to vevs 
at far lower mass scales)
\bea
0&=& \frac{9|f|^2}{m_P^2}|v_X^2|^2v_Y +\frac{f^*A_f^*}{m_P}v_X^{*3}+m_Y^2 v_Y\\
0&=& \frac{3|f|^2}{m_P^2}|v_X^2|^2v_X +\frac{18|f|^2}{m_P^2}|v_X|^2 |v_Y|^2v_X 
+\frac{3 f^* A_f^*}{m_P}v_X^{*2}v_Y^*+m_X^2 v_X .
\eea
To simplify, we will take $A_f$ and $f$ to be real so that the vevs $v_X$ 
and $v_Y$ are real as well. 
Then, the first of these may be solved for $v_Y$ and substituted
into the second equation to yield a cubic polynomial in $v_X^4$ which can 
be solved for either analytically or numerically. 
Viable solutions can be found for $|A_f|\ge \sqrt{12} m_0\simeq 3.46 m_0$ 
(where for simplicity, we assume a common scalar mass $m_X=m_Y= m_{3/2}\equiv m_0$).
Then, for typical soft terms of order $m_{soft}\sim 10$ TeV and $f=1$, we 
develop vevs $v_X\sim v_Y\sim 10^{11}$ GeV. 
For instance, for $m_X=m_Y=10$ TeV, $f=1$ and $A_f=-35.5$ TeV, 
then $v_X=10^{11}$ GeV, $v_Y=5.8\times 10^{10}$ GeV, 
$v_{PQ}\equiv\sqrt{v_X^2+v_Y^2}=1.15\times 10^{11}$ GeV and the PQ scale 
$f_a=\sqrt{v_X^2+9v_Y^2}=2\times 10^{11}$ GeV. 
The $\mu$ parameter for $\lambda_{\mu}=0.1$ is given as 
$\mu= \lambda_\mu v_X^2/m_P\simeq 417$ GeV.

\begin{figure}[tbp]
\includegraphics[height=0.4\textheight]{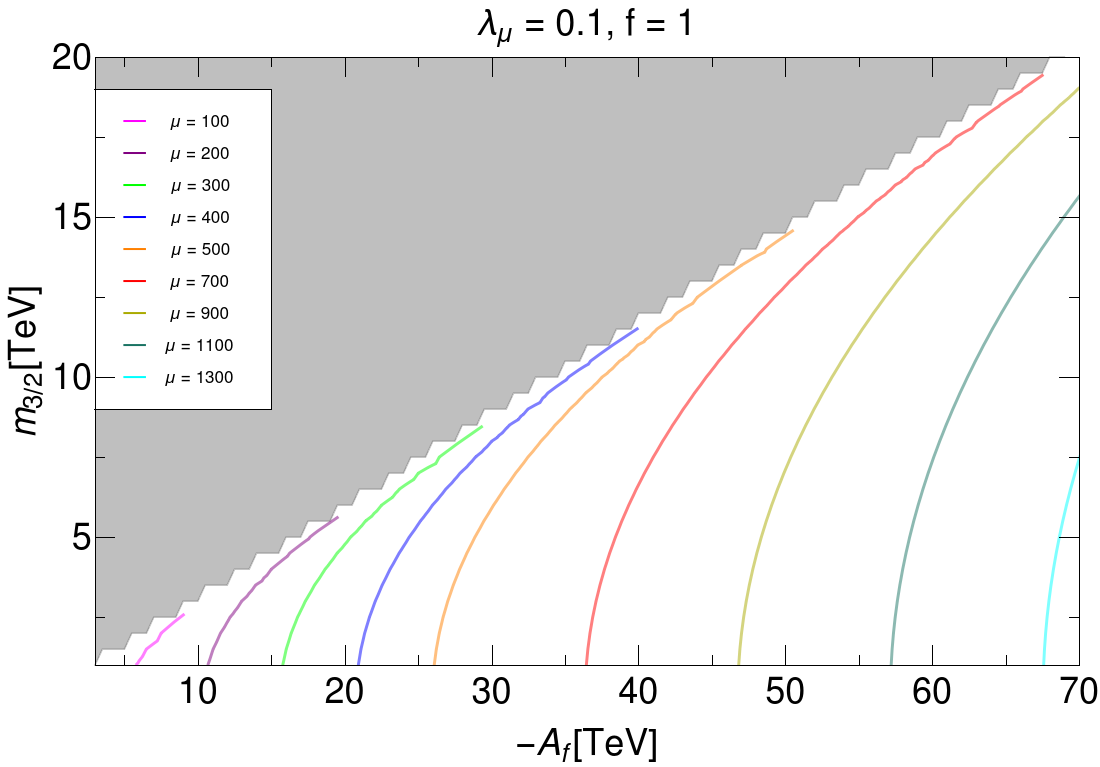}
\caption{Calculated value of SUSY $\mu$ parameter from the
GSPQ model in the $m_{3/2}$ vs. $-A_f$ plane for $f=1$ and
$\lambda_\mu =0.1$.
\label{fig:mu}}
\end{figure}

In Fig. \ref{fig:mu}, we plot contours of the derived value of 
$\mu$ in the $m_{3/2}$ vs. $-A_f$ parameter space for $\lambda_\mu =0.1$. 
The gray-shaded region does not yield admissible vacuum solutions while 
the right-hand region obeys the above bound $|-A_f|\agt\sqrt{12} m_{3/2}$.
From the plot we see that, for any fixed value of gravitino mass $m_{3/2}$, 
low values of $\mu$ occur for the lower allowed range of  $|A_f|$.
There is even a tiny region with $\mu <100$ GeV in the lower-left which may be 
ruled out by negative search results for pair production of 
higgsino-like charginos at LEP2. 
As $|A_f|$ increases, then the derived value of $\mu$ increases beyond 
the anthropic limit of $\mu\alt 350$ GeV and would likely lead to 
too large a value of the weak scale unless an unnatural finetuning 
is invoked in $m_{Z}^{PU}$.

\subsection{GSPQ model in the multiverse}

To begin our calculation of the expected distribution of 
the $\mu$ parameter from the landscape, we adopt the 
two-extra-parameter non-universal Higgs SUSY model 
NUHM2\cite{nuhm2,nuhm22,nuhm23,nuhm24,nuhm25,nuhm26} 
where matter scalar soft masses are
unified to $m_0$ whilst Higgs soft masses $m_{H_u}$ and $m_{H_d}$ 
are independent.\footnote{It is more realistic to allow independent 
generations $m_0(1)$, $m_0(2)$ and $m_0(3)$ but these will hardly affect our results here. They do play a big role in a landscape solution to the 
SUSY flavor and $CP$ problems where $m_0(1)$ and $m_0(2)$ are drawn to 
common upper bounds in the $20-50$ TeV range leading to a mixed
decoupling/quasi-degeneracy solution to the aforementioned problems.} 
The latter soft Higgs masses are usually traded for
weak scale parameters $\mu$ and $m_A$ so the parameter space is given by
\be
m_0,\ m_{1/2},\ A_0,\ \tan\beta,\ \mu,\ m_A\ \ \ \ (NUHM2).
\ee 
We will scan soft SUSY breaking terms with the $n=1$ landscape power-law
draw, with an independent draw for each category of soft term\cite{Baer:2020vad}. 
The scan must be made with parameter space limits beyond those which are 
anthropically imposed. Our p-space limits are given by
\bea
m_0&:&\ 0.1-20\ TeV,\\
m_{1/2}&:&\ 0.5-5\ TeV,\\
-A_0&:&\ 0-50\ TeV,\\
m_A&:&\ 0.3-10\ TeV,\\
\tan\beta &:&\ 3-60\ \ \ (uniform\ scan) 
\eea 
A crucial assumption is that the matter scalar masses in the PQ sector
are universal with the matter scalar masses in the visible sector: 
hence, we adopt that $m_0=m_X=m_Y\equiv m_{3/2}$. 
We also assume correlated trilinear soft terms: $A_f=2.5 A_0$. 
This latter requirement is forced upon us by requiring $|A_f|\ge\sqrt{12}m_0$
to gain a solution in the PQ scalar potential while in the MSSM sector
if $|A_0|$ is too large, then top squark soft-squared  masses are driven
tachyonic leading to CCB vacua. We also adopt $f=1$ throughout.

For our anthropic requirement, we will adopt the atomic principle
from Agrawal {\it et al.}\cite{Agrawal:1997gf2} 
where $m_{weak}^{PU}\alt (2-5)m_{weak}^{OU}$.
To be specific, we will require $m_Z^{PU}<4m_Z^{OU}$ (which corresponds to
the finetuning measure $\Delta_{EW}<30$\cite{ltr,rns}). 
The finetuned solutions are possible but occur rarely compared to non-finetuned
solutions in the landscape\cite{Baer:2019cae}.
The anthropic requirement results in upper bounds on soft terms 
such as to maintain a pocket-universe weak scale value not-too-displaced 
from its measured value in our universe. 
We also must require no charge-or-color-breaking (CCB)
minima and also an appropriate breakdown in electroweak symmetry ({\it i.e.}
that $m_{H_u}^2$ is actually driven negative such that EW symmetry 
is indeed broken).
Given this procedure, then the value of $\mu$ can be calculated from the
GSPQ model scalar potential minimization conditions and then the entire
SUSY spectrum can be calculated using the Isajet/Isasugra package\cite{isajet}.
The resulting spectra can then be accepted or rejected according to the 
above anthropic requirements.

\subsubsection{Results for GSPQ model with $\lambda_{\mu} =0.1$}

\begin{figure}[tbp]
\includegraphics[height=0.3\textheight]{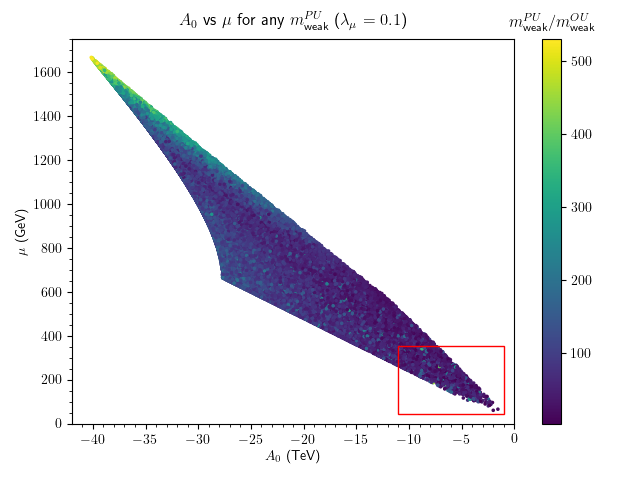}
\includegraphics[height=0.3\textheight]{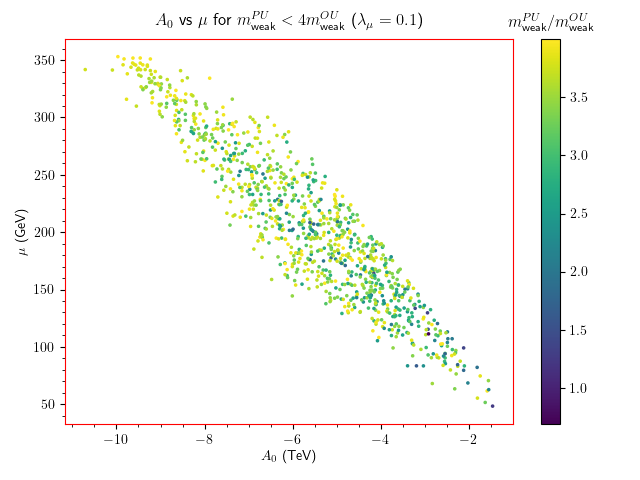}
\caption{Locus of $n=1$ landscape scan points in the
GSPQ+MSSM model in the $A_0$ vs. $\mu$ plane for {\it a}) 
all values of $m_{weak}^{PU}$ and {\it b}) points with 
$m_{weak}^{PU}<4 m_{weak}^{OU}$. We take $f=1$ and $\lambda_\mu =0.1$.
\label{fig:A0vsmu}}
\end{figure}

\begin{figure}[tbp]
\includegraphics[height=0.4\textheight]{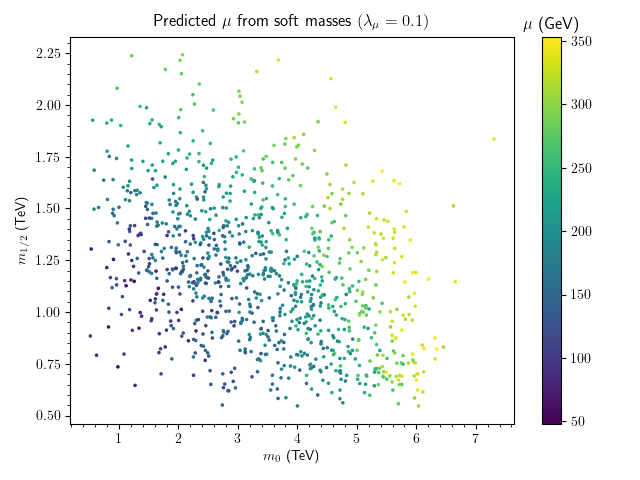}
\caption{Locus of $n=1$ landscape scan points in the
GSPQ+MSSM model in the $m_0$ vs. $m_{1/2}$ plane for 
points with $m_{weak}^{PU}<4 m_{weak}^{OU}$. 
The color coding follows the magnitude of the $\mu$ parameter.
We take $f=1$ and $\lambda_\mu =0.1$.
\label{fig:m0mhf}}
\end{figure}

In this subsection, we restrict our results to parameter scans with
$\lambda_\mu =0.1$. 
In Fig. \ref{fig:A0vsmu}, we show the distribution of scan points in the 
$A_0$ vs. $\mu$ plane for {\it a}) all derived weak scale values  
$m_{weak}^{PU}$ and {\it b}) for only points with $m_{weak}^{PU}<4 m_{weak}^{OU}$.
From frame {\it a}), we see that only the colored portion of parameter space
yields appropriate EWSB, albeit mostly with a huge value of 
$m_{weak}^{PU}$ well beyond the ABDS anthropic window. The points 
with too low a value of $-A_0$ do not yield viable GSPQ vacua
(unless compensated for with an appropriately small value of $m_0$) 
while points with too large a value of $-A_0$ typically yield CCB minima in the
MSSM scalar potential. The surviving points are color coded according
to the value of $m_{weak}^{PU}$ with the dark blue points yielding the lowest
values of $m_{weak}^{PU}$, which occur in the lower-right corner.
In frame {\it b})-- which is a blow-up of the red-bounded region from frame {\it a})-- we add the anthropic condition $m_{weak}^{PU}<4 m_{weak}^{OU}$.
In this case, the range of $-A_0$ and $\mu$ values becomes greatly restricted
since the large $\mu$ points require large values of $m_0$ and $m_{1/2}$, 
leading to too large values of $\Sigma_u^u(\tst_{1,2})$.
This can be seen from Fig. \ref{fig:m0mhf}, where we plot the 
color-coded $\mu$ values in the $m_0$ vs. $m_{1/2}$ plane for
$\lambda_{\mu}=0.1$. From the right-hand scale, the dark purple dots have 
$\mu\alt 100$ GeV (and so would be excluded by LEP2 chargino pair searches
which require $\mu \agt 100$ GeV). The green and yellow points all have large values of $\mu\sim 300-350$ GeV, but these occur at the largest 
values of $m_0$ and $m_{1/2}$. For even larger $m_0$ and $m_{1/2}$ values, 
the derived $\mu$ value exceeds $365$ GeV; 
and absent fine-tuning, such points would lead to $m_{weak}^{PU}$ 
lying beyond the ABDS window, in violation of the atomic principle.

\begin{figure}[tbp]
\includegraphics[height=0.4\textheight]{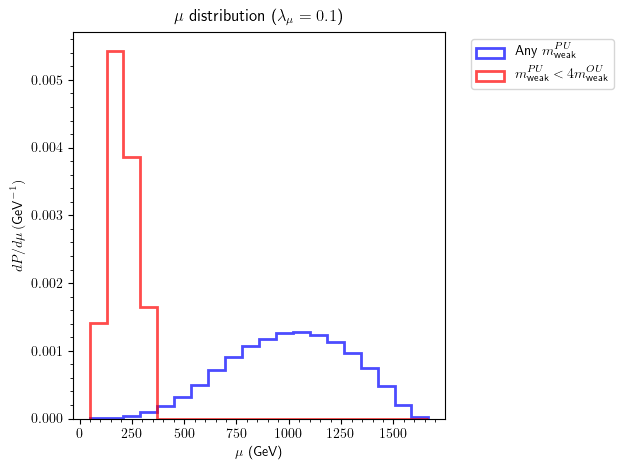}
\caption{Probability distribution for SUSY $\mu$ parameter in the 
GSPQ+MSSM model from an $n=1$ landscape draw to large soft terms 
with $f=1$ and $\lambda_\mu =0.1$.
\label{fig:mu_dist_pt1}}
\end{figure}

In Fig. \ref{fig:mu_dist_pt1}, we plot the distribution of 
derived values of $\mu$ for the GSPQ+NUHM2 model for all
values of $m_{weak}^{PU}$ (blue histogram) and for the 
anthropically-limited points with $m_{weak}^{PU}<4 m_{weak}^{OU}$ (red histogram).
We see the blue histogram prefers huge values of $\mu$, and only turns over 
at high values due to the artificial upper limits we have placed on our 
soft term scan values. However, once the anthropic constraint is applied, 
then we obtain the red distribution which varies between $\mu\sim 50-365$ GeV
with a peak at $\mu^{PU}\sim 200$ GeV followed by a fall-off to larger values.

\begin{figure}[tbp]
\includegraphics[height=0.4\textheight]{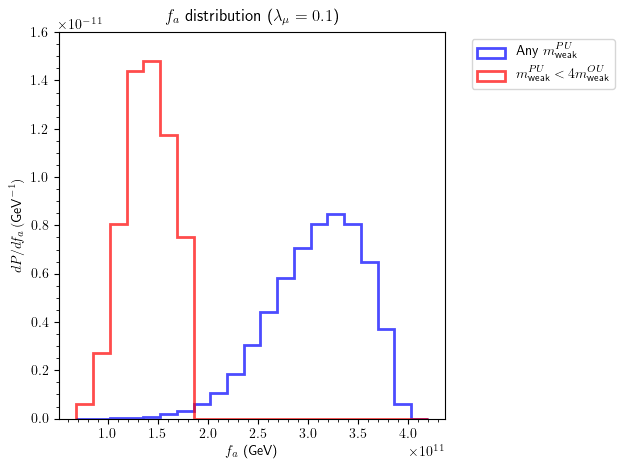}
\caption{Probability distribution for PQ scale $f_a$ in the 
GSPQ+MSSM model from an $n=1$ landscape draw to large soft terms 
with $f=1$ and $\lambda_\mu =0.1$.
\label{fig:fa_dist_pt1}}
\end{figure}

In Fig. \ref{fig:fa_dist_pt1}, we plot the derived value of the PQ scale
$f_a$ from all models with appropriate EWSB (blue) and those models with
$m_{weak}^{PU}<4 m_{weak}^{OU}$ (red). In this case, the PQ scale comes out 
in the cosmological sweet spot where there are comparable relic abundances
of SUSY DFSZ axions and higgsino-like WIMP dark matter\cite{Baer:2019uom}.
The unrestricted histogram ranges up to values of $f_a\sim (2-4)\times 10^{11}$
GeV. This differs from an earlier work which sought to derive the PQ scale
from the landscape by imposing anthropic conditions using constraints 
on an overabundance of mixed axion-neutralino dark matter\cite{Baer:2019uom}.
In the present case, the GSPQ soft terms are correlated with the 
visible sector soft terms and the latter are restricted by requiring the
derived weak scale to lie within the ABDS window. 
The fact that the present results lie within the cosmological sweet zone 
then resolves a string theory quandary as to why the PQ scale 
isn't up around the GUT/Planck scale\cite{Svrcek:2006yi}.
By including the weak scale ABDS anthropic requirement, the red histogram 
becomes rather tightly restricted to lie in the range 
$f_a:(1-2)\times 10^{11}$ GeV.

\begin{figure}[tbp]
\includegraphics[height=0.4\textheight]{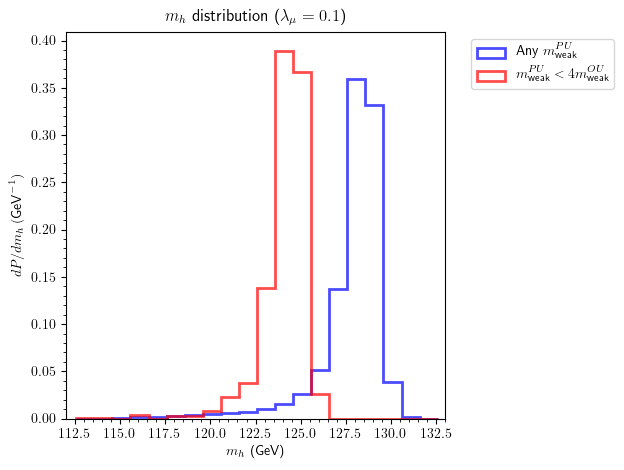}
\caption{Probability distribution for $m_h$ in the 
GSPQ+MSSM model from an $n=1$ landscape draw to large soft terms 
with $f=1$ and $\lambda_\mu =0.1$.
\label{fig:mh_dist_pt1}}
\end{figure}

In Fig. \ref{fig:mh_dist_pt1}, we show the expected distribution in light 
Higgs mass $m_h$ without (blue) and with (red) the anthropic constraint. 
For the blue histogram, the upper bound on soft terms is set by a 
combination of our scan limits but also the requirement of getting 
an appropriate breakdown of PQ symmetry (as in lying outside the gray-shaded 
region of Fig.~\ref{fig:mu}). In this case, the distribution peaks 
around $m_h\sim 128$ GeV with only small probability down to
$m_h\sim 125$ GeV. When the anthropic constraint $m_{weak}^{PU}<4m_{weak}^{OU}$
is imposed, then we gain instead the red histogram which features a
prominent peak around $m_h\sim 125$ GeV, which is supported 
by the ATLAS/CMS measured value of $m_h$\cite{Zyla:2020zbs}.

\begin{figure}[tbp]
\includegraphics[height=0.4\textheight]{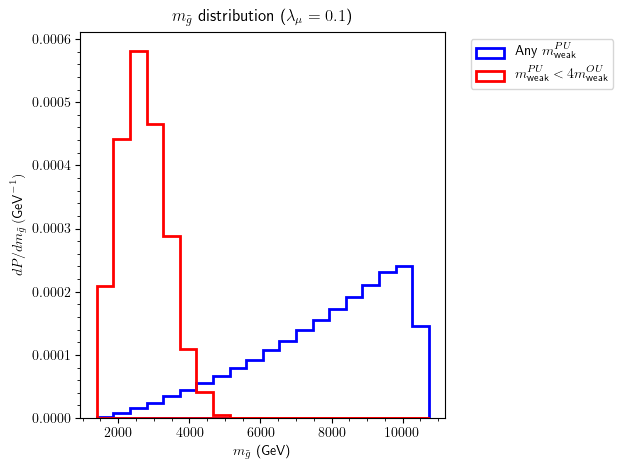}
\caption{Probability distribution for $m_{\tg}$ in the 
GSPQ+MSSM model from an $n=1$ landscape draw to large soft terms 
with $f=1$ and $\lambda_\mu =0.1$.
\label{fig:mg_dist_pt1}}
\end{figure}

In Fig. \ref{fig:mg_dist_pt1}, we show the expected distribution in gluino mass 
$m_{\tg}$. For the blue curve, without the anthropic constraint, 
we have a strong statistical draw from the landscape for large gluino 
masses which is only cut off by our artificial upper scan limits along with
the requirement of appropriate PQ breaking. Once the anthropic condition 
is imposed, then the $m_{\tg}$ distribution peaks around $m_{\tg}\sim 3$ TeV
with a tail extending up to about 5 TeV. 
The ATLAS/CMS requirement that $m_{\tg}\agt 2.2$ TeV only restricts the 
lowest portion of the derived $m_{\tg}$ probability distribution.

\subsubsection{Results for other values of $\lambda_\mu$}

\begin{figure}[tbp]
\includegraphics[height=0.4\textheight]{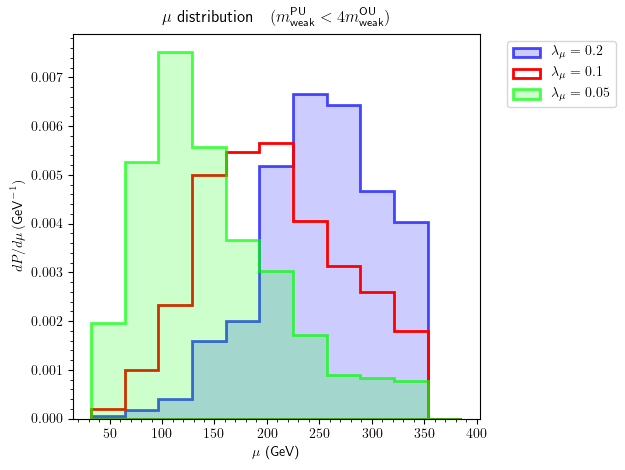}
\caption{Probability distribution for SUSY $\mu$ parameter in the 
GSPQ+MSSM model from an $n=1$ landscape draw to large soft terms 
with $f=1$ for $\lambda_\mu = 0.05,\ 0.1$ and $0.2$.
\label{fig:mu_dist_comp}}
\end{figure}

We have repeated our calculations to include other choices of 
$\lambda_\mu=0.02,\ 0.05,\ 0.1$ and $0.2$. By lowering the value of
$\lambda_\mu$, then correspondingly larger GSPQ soft term values 
(and hence NUHM2 soft term values) may lead to acceptable vacua.
In Fig. \ref{fig:mu_dist_comp}, we show the derived $\mu$ parameter
distribution for three choices of $\lambda_\mu$ after the 
anthropic weak scale condition is applied. 
A fourth histogram for $\lambda_\mu =0.02$ actually peaks 
below $\sim 100$ GeV and so the bulk of this distribution would be 
ruled out by LEP2 limits which require $\mu\agt 100$ GeV due to
negative searches for chargino pair production. As $\lambda_\mu$ increases,
then the $\mu$ distribution becomes correspondingly harder: for 
$\lambda_\mu=0.2$, then the distribution actually peaks around
$\mu\sim 250-300$ GeV. This could offer an explanation as to why
ATLAS and CMS have not yet seen the soft dilepton plus jets plus $\eslt$
signature which arises from higgsino 
pair production\cite{Baer:2011ec,Han:2014kaa,Baer:2014kya,Han:2015lma,Baer:2020sgm} 
at LHC\cite{Aad:2019qnd,CMS:2021xji}. 
Current limits on this process from ATLAS extend out to $\mu\sim 200$ GeV 
for $m_{\tchi_2^0}-m_{\tchi_1^0}$ mass gaps of $\sim 10$ GeV\cite{Aad:2019qnd,CMS:2021xji}.

\begin{figure}[tbp]
\includegraphics[height=0.4\textheight]{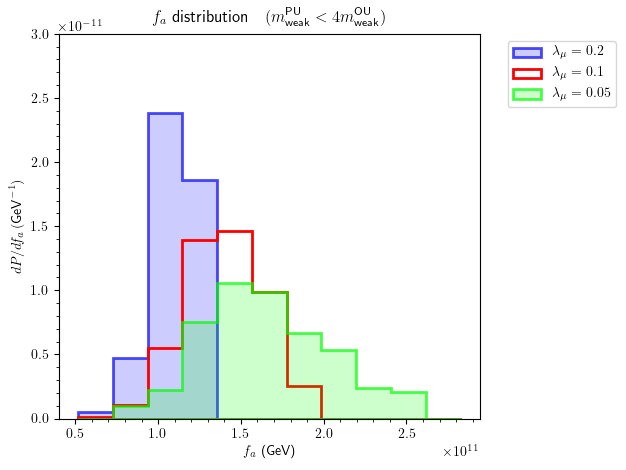}
\caption{Probability distribution for PQ scale $f_a$ in the 
GSPQ+MSSM model from an $n=1$ landscape draw to large soft terms 
with $f=1$ and $\lambda_\mu =0.05,\ 0.1$ and $0.2$.
\label{fig:fa_dist_comp}}
\end{figure}

In Fig. \ref{fig:fa_dist_comp}, we show the distribution in $f_a$ for 
the three different values of $\lambda_\mu$. Here the model is rather
predictive with the PQ scale lying at $f_a\sim (0.5-2.5)\times 10^{-11}$
GeV, corresponding to an axion mass of $m_a\sim 144-720$ $\mu$eV.
Unfortunately, in the PQMSSM, the axion coupling $g_{a\gamma\gamma}$ is 
highly suppressed compared to the non-SUSY DFSZ model due to 
cancelling contributions from higgsino states circulating in 
the $a\gamma\gamma$ axion coupling triangle diagram\cite{Bae:2017hlp}. 
Thus, axion detection at experiments like ADMX may require 
new advances in sensitivity in order to eek out a signal.  

\section{Distribution of $\mu$ parameter in Giudice-Masiero model}
\label{sec:GM}

For GM, one assumes first that the $\mu$ parameter is 
forbidden by some symmetry ($R$-symmetry or Peccei-Quinn (PQ) symmetry?).
Then one assumes that in the SUSY K\"ahler potential $K$, 
there is a Planck suppressed coupling of the Higgs bilinear to
some hidden sector field $h_m$ which gains a SUSY-breaking vev:
\be
K_{GM}\ni \lambda_{GM} h_m^\dagger H_u H_d/m_{P} +c.c.
\ee
where $\lambda_{GM}$ is some Yukawa couping of order $\sim 1$. 
When $h_m$ develops a SUSY breaking vev $F_h\sim m_{hidden}^2$ with 
the hidden sector mass scale $m_{hidden}\sim 10^{11}$ GeV, then a 
weak scale value of 
\be
\mu_{GM} \simeq \lambda_{GM} m_{hidden}^2/m_P
\ee
 would ensue, where $m_P$ is the reduced Planck mass 
$m_P=m_{Pl}/\sqrt{8\pi }\simeq 2.4\times 10^{18}$ GeV. 
In the GM model, since $\mu\propto F_h$ (a single $F$-term), 
then one would expect also that $\mu_{GM}$ would scale as $m_{soft}^1$ 
in the landscape. Nowadays, models invoking the $\mu$-forbidding PQ
global symmetry are expected to lie within the swampland of
string-inconsistent theories since quantum gravity admits no global
symmetries\cite{Banks:1988yz,Kallosh:1995hi,Daus:2020vtf}. 
Discrete or continuous $R$-symmetries or gauge symmetries 
may still be acceptable; the former are expected to emerge from 
compactification of manifolds with higher dimensional spacetime symmetries.

\begin{figure}[tbp]
\includegraphics[height=0.4\textheight]{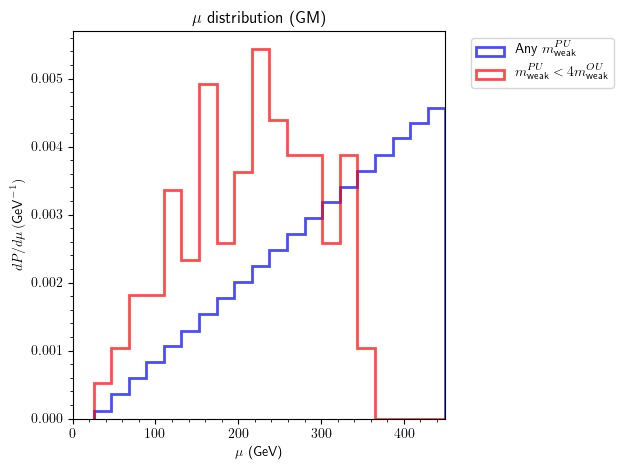}
\caption{Distribution of SUSY $\mu$ parameter in the GM model
with $\lambda_{GM} =1$ with and without the anthropic constraint
that $m_{weak}^{PU}<4 m_{weak}^{OU}$.
\label{fig:muGM}}
\end{figure}

In Fig. \ref{fig:muGM}, we show the expected distribution of the 
$\mu_{GM}$ parameter ($\mu$ in the GM model) without (blue) and with (red)
the anthropic constraint that $m_{weak}^{PU}<4 m_{weak}^{OU}$.
The blue histogram is just a linear expectation of the $\mu$ 
parameter up to the upper scan limit. Thus, for the GM model in the landscape, 
one expects a huge $\mu$ parameter. 
Varying the coupling $\lambda_{GM}$ just rescales the $\mu_{GM}$ 
distribution. And since the $\mu_{GM}$ sector effectively
decouples from the visible sector (unlike for the GSPQ model), 
we do not find that varying $\lambda_{GM}$ has any effect 
on the expected $\mu_{GM}$ distribution from the landscape.

Next, the $\mu_{GM}$ distribution must be tempered by the 
anthropic constraint which then places an 
upper limit of $\mu\alt 365$ GeV, but also excludes some parameter
space with too large $\Sigma_u^u$ values. 
Here, for $\lambda_{GM}=1$, we see the expected $\mu$ parameter 
distribution peaks around $\sim 250$ followed by a drop-off to $\sim 360$ GeV.

\section{Summary and conclusions}
\label{sec:conclude}

In this paper we have explored the origin of several mass scale mysteries
within the MSSM as expected from the string landscape. 
Soft SUSY breaking terms are expected to be distributed as a power-law 
or log distribution (although in dynamical SUSY breaking they are
expected to scale as $1/m_{soft}$\cite{Baer:2021uxe}). 
But other mass scales arise in supersymmetric models: 
the SUSY conserving $\mu$ parameter, the PQ scale $f_a$ 
(if a solution to the strong CP problem is to be included)
and the Majorana neutrino scale $M_N$. 
Here, we have examined the expected
distribution of the SUSY $\mu$ parameter from the well-motivated
GSPQ model which invokes a discrete $\mathbb{Z}_{24}^R$ symmetry to forbid
the $\mu$ term (along with $R$-parity violating terms and while
suppressing dangerous $p$-decay operators). 
It also generates an accidental, approximate global PQ symmetry
which is strong enough to allow for the theta parameter 
$\bar{\theta}\alt 10^{-10}$ (hence it is gravity-safe\cite{PQgrav1,PQgrav2,PQgrav3,PQgrav4}).
The breaking of SUSY in the PQ sector then generates 
a weak scale value for the $\mu$ parameter and
generates a gravity-safe PQ solution to the strong CP problem.
For the GSPQ model, we expect the PQ sector soft terms to be 
correlated with visible sector soft terms which scan on the landscape
and are susceptible to the anthropic condition that 
$m_{weak}^{PU}<4m_{weak}^{OU}$ in accord with the ABDS window. 
Thus, a landscape distribution for both the $\mu$ parameter and 
the PQ scale $f_a$ are generated. For small values of Yukawa coupling 
$\lambda_\mu$, then the $\mu$ distribution is stilted towards low values
$\mu\sim 100$ GeV which now seems ruled out by recent ATLAS/CMS
searches for the soft-dilepton plus jets plus $\eslt$ signature
which arises from light higgsino pair production at LHC. 
For larger values of $\lambda_\mu\sim 0.1-0.2$, then the $\mu$ 
distribution is stilted towards large values $\mu\sim 200-300$ GeV
in accord with LHC constraints. The PQ scale $f_a$ also ends up lying
in the cosmological sweet zone so that dark matter would be comprised
of an axion/higgsino-like WIMP admixture\cite{Bae:2013bva,Bae:2013hma,Bae:2014rfa,Bae:2017hlp}. 

We also examined the $\mu$ distribution expected from the Giudice-Masiero 
solution. In this case, the $\mu$ parameter is expected to scan 
as $m_{soft}^1$ with a distribution peaking around $\mu\sim 200-300$ GeV.

\section*{Acknowledgments}

This material is based upon work supported by the U.S. Department of Energy, 
Office of Science, Office of High Energy Physics under 
Award Number DE-SC-0009956 and U.S. Department of Energy (DoE) Grant DE-SC-0017647. 
The computing for this project was performed at the OU Supercomputing Center 
for Education \& Research (OSCER) at the University of Oklahoma (OU).


\bibliography{mu3}
\bibliographystyle{elsarticle-num}

\end{document}